\documentclass[aps,pra,reprint,superscriptaddress]{revtex4-1}
\usepackage{graphicx}
\usepackage{amsmath}

\draft % marks overfull lines with a black rule on the right
 \usepackage{bm}
 \usepackage{mathrsfs}
\usepackage{float}
\begin{document}

%\preprint{}
%COPS INTERNAL REPORT SS141010      


\title{Measurement of the profiles of disorder-induced localized resonances in photonic crystal waveguides by local tuning}

\author{Jin Lian$^{1,2,*}$, Sergei Sokolov$^{1,2}$, Emre Y\"uce $^{2,3}$, Sylvain Combri\'e$^{4}$,  Alfredo De Rossi$^{4}$, and Allard P. Mosk$^{1}$ }

\affiliation{
$^1$Nanophotonics, Debye Institute for Nanomaterials Science, Center for Extreme Matter and Emergent Phenomena, Utrecht University, P.O. Box 80.000, \\3508 TA  Utrecht, The Netherlands \\
$^2$Complex Photonic Systems (COPS), MESA+ Institute for
Nanotechnology, University of Twente, P.O. Box 217, \\ 7500AE  Enschede, The Netherlands \\
$^3$Light $\&$ Matter Control Group, Department of Physics,Middle East Technical University, \\ 06800 Ankara, Turkey\\
$^4$Thales Research and Technology, Route Départementale 128, \\91767 Palaiseau, France\\
$^*$Corresponding author: j.lian@uu.nl
}

%\thanks{}
\makeatletter
\let\setrefcountdefault\relax
\makeatother
%\date{SS141010 \today}
\date{\today}

\begin{abstract}
Near the band edge of photonic crystal waveguides, localized modes appear due to disorder. We demonstrate a new method to elucidate spatial profile of the localized modes in such systems using precise local tuning. Using deconvolution with the known thermal profile, the spatial profile of a localized mode with quality factor ($Q$) $>10^5$ is successfully reconstructed with a resolution of $2.5 \ \mu $m. 
\end{abstract}


\maketitle %\maketitle must follow title, authors, abstract and \pacs

%\section{}
 %\label{}
%\subsection{}
%\subsubsection{}
\noindent 
\section{Introduction}
\begin{figure}[H]
\vspace*{0.0in}
 %vspace{0.5cm}
\centerline{\includegraphics[width=0.9\columnwidth]{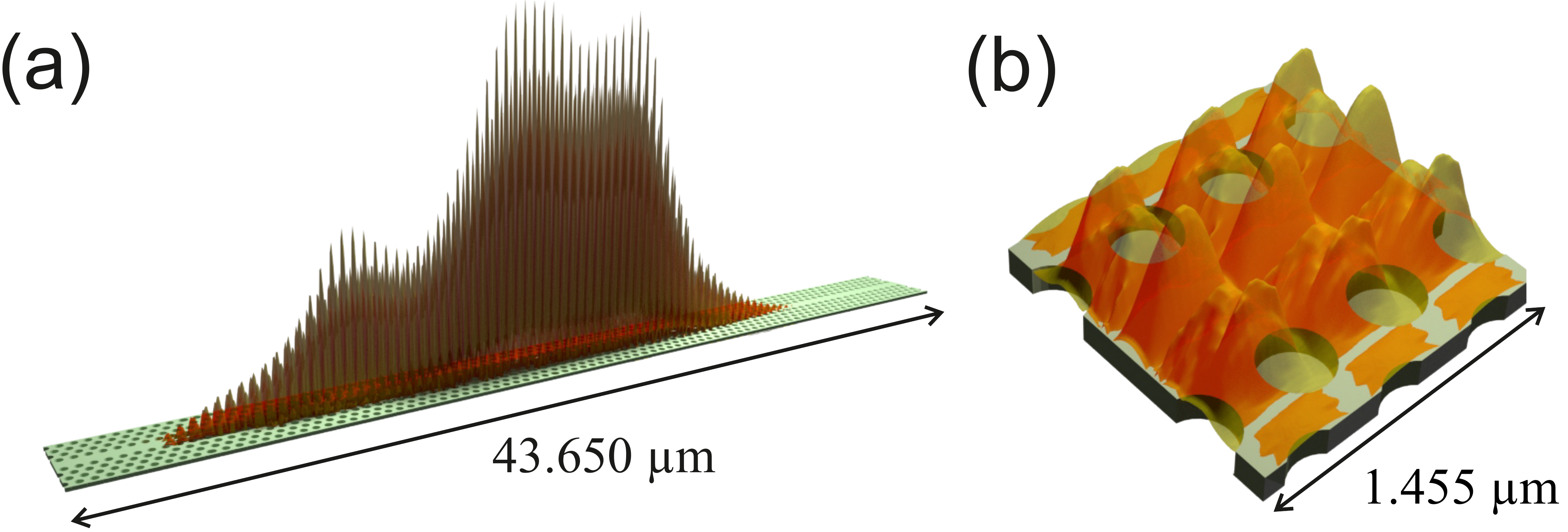}}
\caption{(a) Visualization of the intensity of a typical disorder induced localized mode in a photonic crystal waveguide. (b) Zooming in of the center part of (a), the intensity is scaled down in order to show that the periodic oscillation of the intensity matches the period of the waveguide.}  
\end{figure}

Disorder-induced multiple scattering of waves is a universal phenomenon in physics. Coherent scattering in a disordered system leads to localization of waves \cite{anderson1956,lagendijk2009}. For a periodical system with disorder, localization of waves appears near the edge of the energy band \cite{Anderson1972,John1984}. In photonics, photonic crystal waveguides (PhCWG) \cite{mitbook} are well-known periodic systems where localization of light  \cite{Patterson2009,mazoyer2009,Savona2011,Spasenovic2012,Huisman2012,Baron2015} is unavoidable near the band edge because of the ubiquitous disorder caused by fabrication \cite{femius2005,Kuramochi2005}.  Localization of light in PhCWGs appears as a set of spatially random localized modes. 

An example of a disorder-induced random localized mode is shown in Fig. 1. The numerically calculated intensity profile shows an envelope that is approximately 30 $\mu$m wide, and oscillates with a carrier frequency that corresponds to the waveguide mode at the edge of the Brillouin zone. Remarkably, the disorder on the 1-nm scale shows no noticeable effect on the local structure of the mode wavefunction, its main effect is the modification of the envelope which is around 10 $\mu$m in scale.  This shows the the most important information of the random localized mode is the spatial profile, the envelope of the mode. 

The random localized modes raise much interest because of their inherently high quality factor ($Q$), which maybe utilized for random quantum networks \cite{Sapienza2010}. The spatial and the spectral characteristics of disorder induced modes in PhCWGs are not predictable due to the disordered nature. In previous works, characterizations have been done using invasive methods that perturb the structure, such as near field scanning optical microscopy (NSOM) \cite{Huisman2012,Spasenovic2012} and enhanced emission measurements from the quantum dots in the PhCWGs \cite{Sapienza2010}. Time dependent optical tuning by carrier injection induces local index shift and absorption \cite{Bruck2014}, hence it is not applicable for high $Q$ systems.  

Here, we show a new non-invasive method to elucidate the spatial profile of the localized modes in a photonic crystal waveguide with unavoidable disorder.  In this method, we firstly use precise local thermal tuning \cite{Sokolov2015} to obtain the resonance wavelength shift as a function of position.  Secondly, using deconvolution with the known thermal profile, we reconstruct the mode profiles of the disorder-induced modes. Using our method, we successfully  reconstruct the spatial profile a localized mode with $Q >10^5 $ with a resolution of 2.5 $\mu$m.

\section{Sample and Apparatus}
\begin{figure}[htp]
\centerline{\includegraphics[width=0.85\columnwidth]{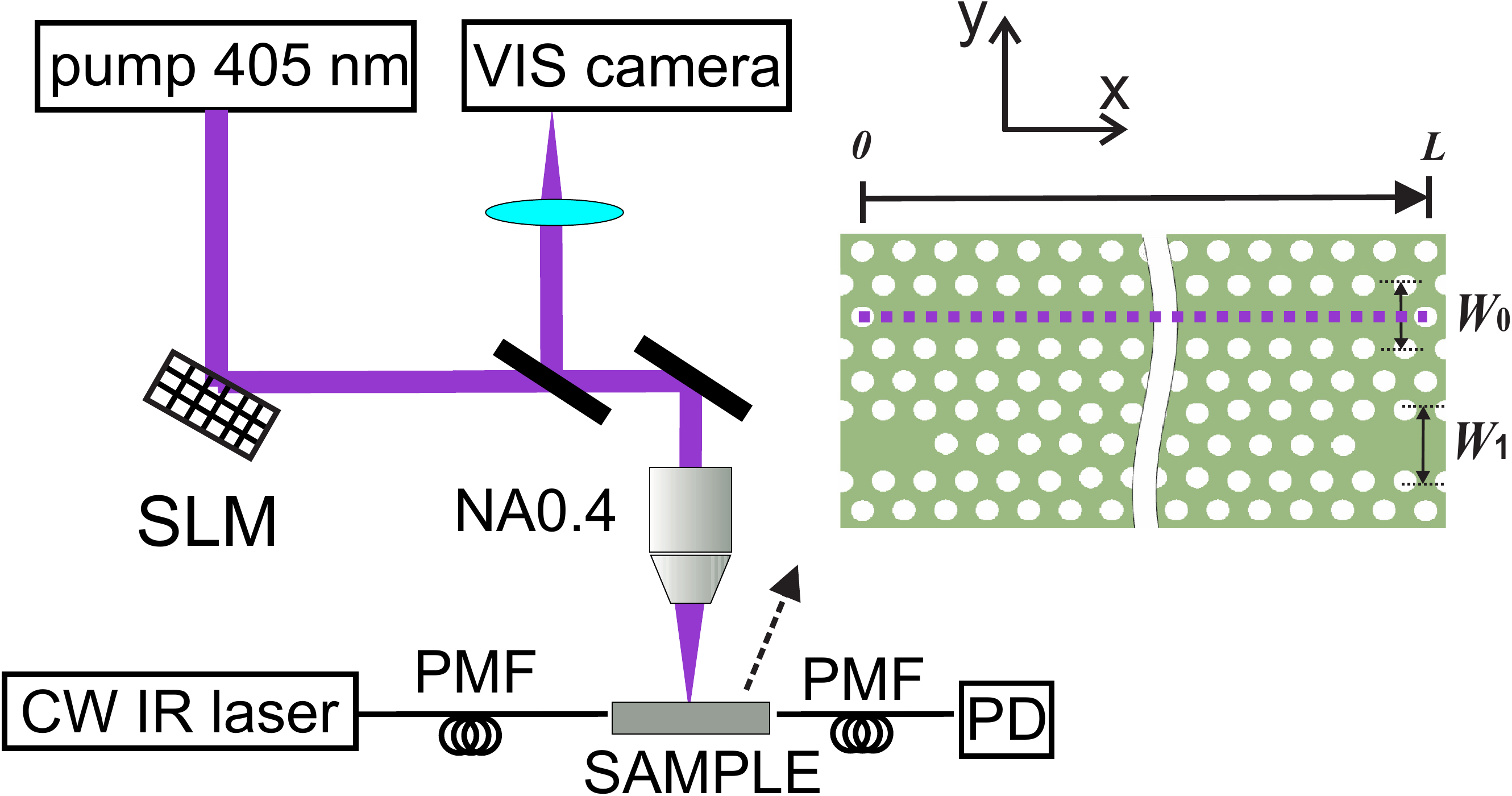}}
\caption{Experimental setup and schematic representation of the sample.  A Photodiode (PD) is used to detect the transmission signal. Polarization maintaining lensed fibers (PMF) are used to couple light from the CW IR laser to the sample and couple light out of sample to the PD. }
\end{figure}
The sample and our experimental setup are shown in Fig. 2. The sample is a triangle lattice photonic crystal membrane structure made of GaInP \cite{alfredoapl2009}. The thickness $h$ of the membrane is 180 nm and the lattice constant $a$ is 485 nm.  A barrier waveguide with width $W_0=0.98\sqrt{3}a$  and length $L=106a$ is created in the sample and it is side coupled to two carrier waveguides with width $W_1=1.1\sqrt{3}a$. The width of $W_1$ guarantees that in the frequency range of interest where disorder-induced resonances show up in the barrier waveguide, the carrier waveguides work in the fast light transport regime in which the influence of disorder is most likely to be only scattering loss rather than forming disorder-induced resonances. A tunable continuous wave (CW) infrared (IR) laser is coupled to the input carrier waveguide of our sample by a polarization maintaining lensed fiber (PMF)  with numerical aperture (NA) of 0.55. The transmitted signal from the sample is collected by another lensed fiber, then detected by a photodiode (PD).  A pump spot from a CW  diode laser ($\lambda_\textrm{pump}=405 \ $nm) is focused on our sample surface by an objective with 0.4 NA. The full width half maximum (FWHM) of the pump spot is 0.83 $\mu$m, and the power is 20 $\mu $W. The surface of the sample is imaged with a visible range camera using a tube lens with a system magnification of $\times27$. We control the position of the pump spot on the barrier waveguide by writing blazed phase gratings with different periods on a spatial light modulator.  To suppress oxidation effects \cite{Caselli2012,Riboli2014}, the sample is kept in a N$_2$ environment with oxygen concentration less than 0.03$\%$ for all measurements.
\section{Transmission spectra}
In Fig. 3(a), we show a reference transmission spectrum of our sample. We see multiple narrow peaks and dips on top of Fabry-P\'erot fringes with Fano line shapes \cite{Miroshnichenko2010}, The narrow resonances correspond to the disorder-induced localized modes in the PhCWG. In Fig. 3(b), we show the transmission spectrum with and without a pump spot at position $x=17.7 \ \mu$m in the waveguide. We observe a pump induced redshift of 163 pm for this resonance. The reference wavelength of this resonance is 1531.515 nm and the observed $Q$ factor for the reference is $3.8 \times 10^5$. Due to the pump laser noise, we see modulation on the intensity of the transmitted signal, however no significant influence on the linewidth is observed.

\begin{figure}[H]%\vspace{0.5cm}
\centerline{\includegraphics[width=.85\columnwidth]{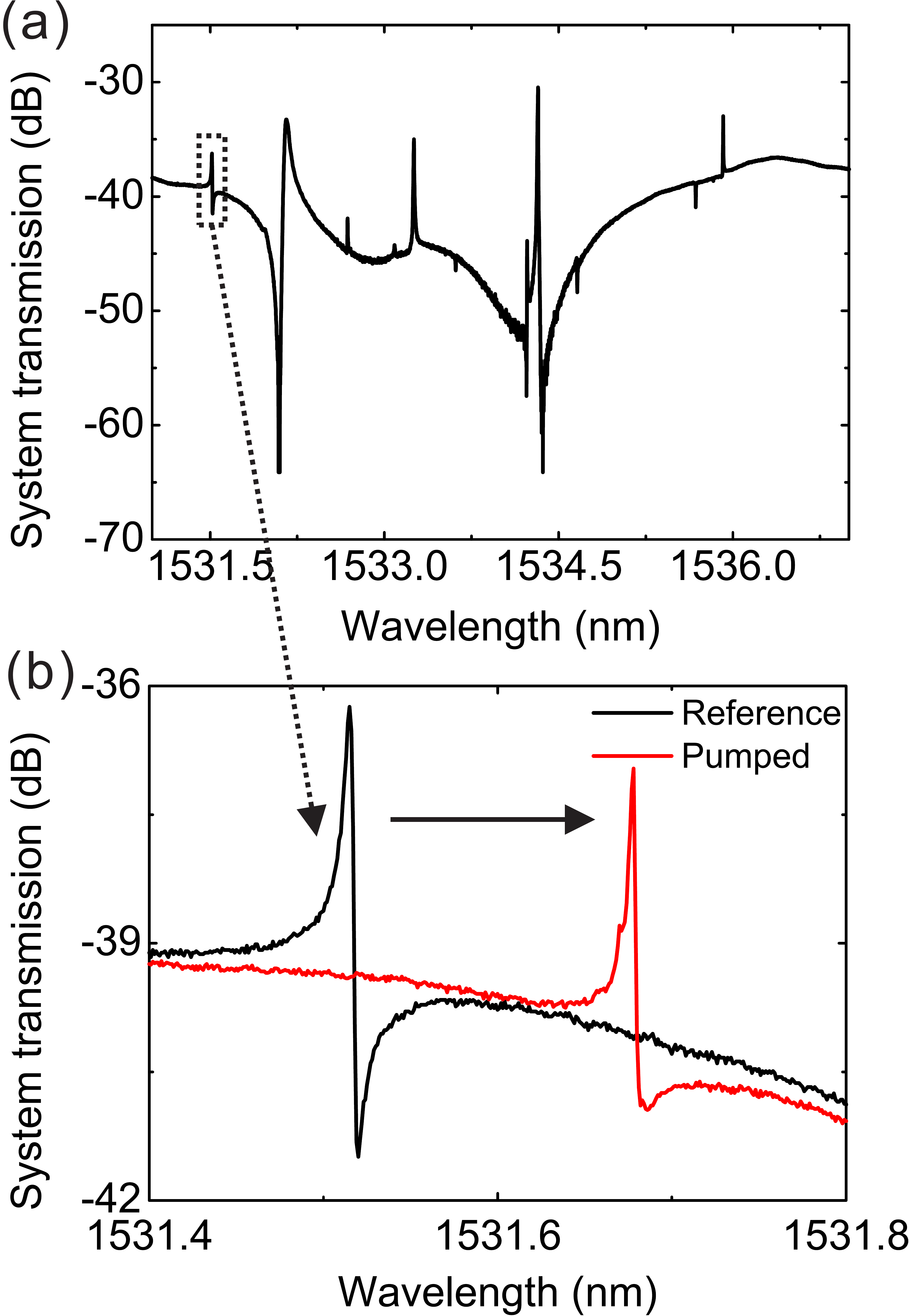}}
\caption{(a) Transmission spectrum of the sample. (b)Zoomed in reference (black) and pumped (red) transmission spectra of the sample. The system transmission here is defined as the ratio between the power we send into the input fiber and the power we detect on the PD. }
\end{figure}

The redshift of the resonance is due to the increase of the dielectric constant of the sample with temperature, and it can be can also be calculated quantitatively by first order perturbation theory \cite{mitbook}, 
\begin{equation}
  \Delta\lambda(\textbf{r}_0)\approx\lambda_0\frac{\int\delta\epsilon(\textbf{r}_0-\textbf{r})|\textbf{E}(\textbf{r})|^2d\textbf{r}}{\int\epsilon(\textbf{r})|\textbf{E}(\textbf{r})|^2d\textbf{r}}.
\end{equation}
Here $\lambda_0$ is the wavelength of the resonance shown in Fig. 3(a). $\textbf{E}(\textbf{r})$ represents the electric field of the mode. $\epsilon(\textbf{r})$ is the dielectric distribution of the waveguide. $\delta\epsilon(\textbf{r}_0-\textbf{r})$ is the perturbation of the dielectric caused by the thermal tuning, it is proportional to the temperature distribution $\delta T(\textbf{r}_0-\textbf{r})$ and $\textbf{r}_0$ is the pump position. In the first order perturbation theory $\textbf{E}(\textbf{r})$ is unchanged for small $\delta\epsilon$. Perturbation theory is only valid when shifts are small compared to the spacing to any nearby mode.
\section{Thermal profile and extracted intensity profile}
From a single transmission spectrum, it is not possible to obtain the spatial information of the modes. With the help of performing the spatial dependent tuning, the spatial information can be retrieved. When we move the pump along the waveguide direction, the wavelength shift we detect is the convolution of the intensity profile of the resonance and the temperature profile that is induced by the pump beam. Taking into account the fact the system is 1D \cite{Henri2010}, the obtained wavelength shift at pump position $x_0$ can be approximated as
%\begin{equation}
%  \Delta\lambda(\textbf{r}_0) = \alpha  {\int\delta T(\textbf{r}_0-\textbf{r})|5\textbf{E}(\textbf{r})|^2d\textbf{r}}.
%\end{equation}, $\alpha$ is a proportional constant. Taking into account the fact the system is quasi 1D, e.q.(2) could be approximated to
\begin{equation}
  \Delta\lambda(x_0) \approx \alpha  {\int\ (T(x_0-x)-T_0)|\textbf{E}^{X}(x)|^2dx}.
\end{equation} Here, $T(x)$ is the temperature profile along the waveguide direction and temperature gradients transversal to the waveguide are neglected, and $T_0$ is the constant offset temperature. $|\textbf{E}^{X}(x)|^2$ is the envelope of the mode profile along the $x$ direction, here $y$ and $z$ dependence of $\textbf{E}(\textbf{r})$ is approximated to be as same as the waveguide mode at the band edge, and can be integrated out. $\alpha$ is a normalization factor.  As a result, the spatial profile of the localized mode can be reconstructed by deconvolving the wavelength shift with temperature distribution. 

We scan our pump beam along the waveguide from $x=0 \ \mu$m to $x=48.7 \ \mu$m. The shift in wavelength of the mode versus pump position is shown in Fig. 4(a).  We see that the wavelength shift varies with the pump position. The maximum shift is observed when the pump is around the middle of the waveguide. This shows the mode is localized in the center part of the waveguide. The temperature distribution caused by the pump beam is calculated using the finite element method (COMSOL) as in Ref. 16. Following the same procedures, we have simulated the temperature distribution in the waveguide, it is shown in Fig. 4(a). We see that although the temperature distribution has a relatively narrow peak, it has a considerably wide base. This indicates that the envelope function of the random localized mode is significantly blurred by the temperature profile.  Thus, in order to reconstruct the mode profile of the disorder-induced localized mode, deconvolution is needed.
\begin{figure}[htb]\vspace{-0.2cm}
\centerline{\includegraphics[width=.85\columnwidth]{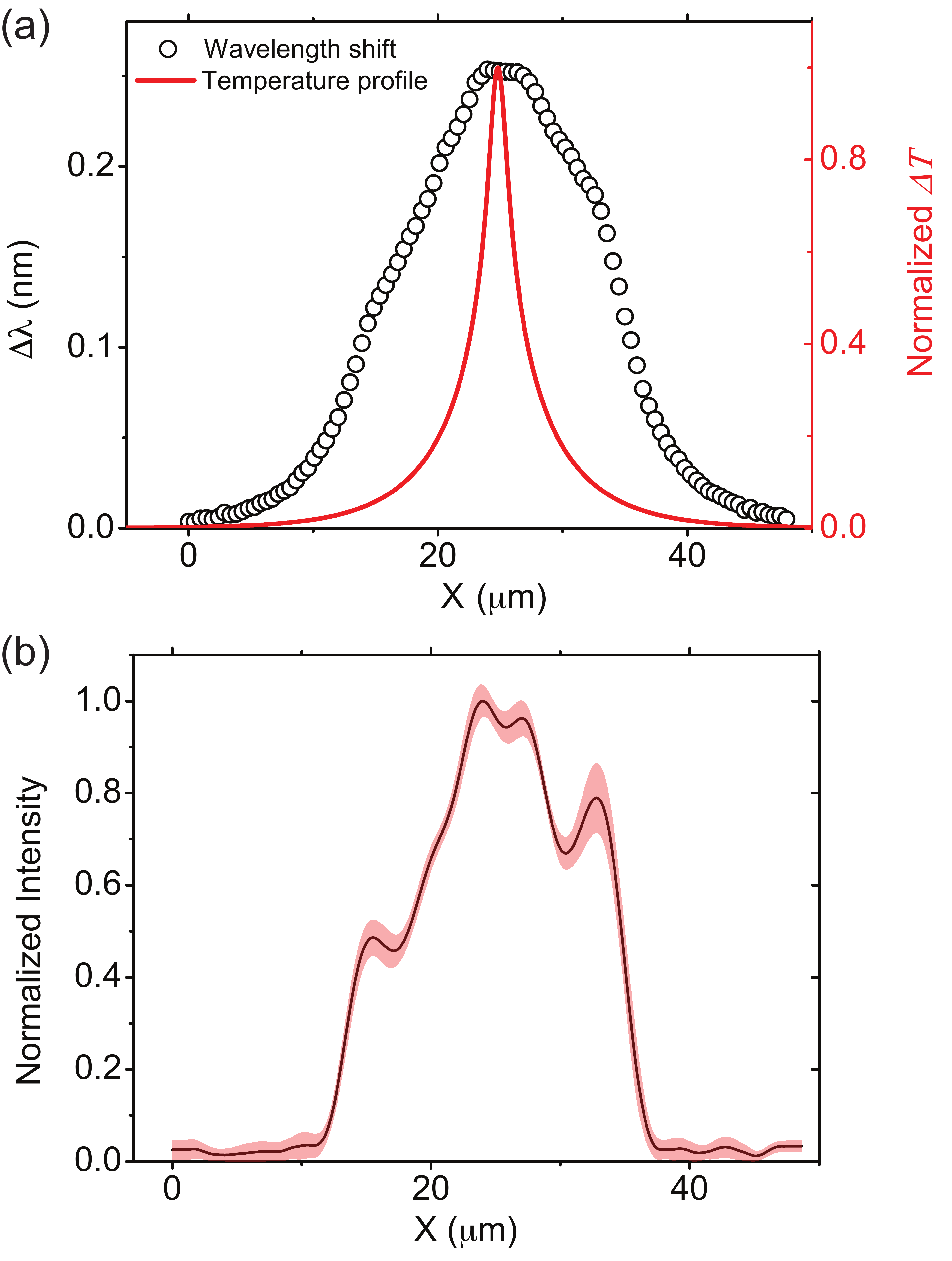}}
\caption{(a) Black circles: mean value of wavelength shift versus pump spot position of 5 independent measurements. The standard deviation is approximately equal to the symbol size. Red curve: temperature distribution along the waveguide direction. (b) Black curve: reconstructed mode profile along the waveguide direction. Pink bar: 3$\sigma _E$ confidence interval.} %\cite{errorbar}.}
%\footnote{The estimated error bar is smoothed $\pm 3 \sigma _E$, and $\sigma _E$ is the standard error of 5 reconstructed spatial profiles}%
\end{figure}

We perform the deconvolution procedures in the Fourier domain. Using the notation $\mathscr{F}$ to represent Fourier transform, we define $I(f_x)=\mathscr{F} \big\{|\textbf{E}^{X}(x)|^2\big\}$, $S(f_x)=\mathscr{F}\big\{\Delta\lambda(x)\big\}$ and $ H (f_x)=\mathscr{F}\big\{ T(x_0-x)-T_0\big\}$. Before the Fourier transform, we interpolate the measured wavelength shift to match the resolution of the simulated temperature profile. $I(f_x)$ is obtained by 
\begin{equation}
  I(f_x) =\frac{S(f_x)}{H (f_x)}G(f_x),
\end{equation}
here $G(f_x)$ is a Butterworth filter. The cutoff of the filter is determined by $\mathrm{SNR}(f_x)$, which in our case is defined as $\mathrm{SNR}(f_x)=\frac{|\overline{S(f_x)}|^2}{\mathrm{Var}[S(f_x)]}$. The cutoff of $G(f_x)$ is selected at the frequency where $\mathrm{SNR}(f_x)=1$. For our experimental data, $f_\textrm{cutoff}=0.4 \ {\mu \textrm{m}}^{-1}$. From the cutoff frequency we estimated our resolution to be 2.5 $\mu$m which is $f_\textrm{cutoff}^{-1}$. $|\textbf{E}^{X}(x)|^2$ can be obtained by $\mathscr{F}^{-1}\big\{I(f_x)\big\}$. 

Using the procedure mentioned above, we deconvolve the wavelength shifts obtained from 5 measurements on the same structure. After the deconvolution, we first calculate the mean value and standard error of the 5 results. Secondly, we estimate the upper and lower confidence limits of the reconstructed signal as $|\textbf{E}^{X}(x)|^2_\pm=\overline{|\textbf{E}^{X}(x)|^2} \pm 3 \sigma_E(|\textbf{E}^{X}(x)|^2)$. $\overline{|\textbf{E}^{X}(x)|^2} $ is the mean of the deconvolved intensities, $\sigma_E(|\textbf{E}^{X}(x)|^2)$ is the standard error. Thirdly, we used simple moving average to smooth $|\textbf{E}^{X}(x)|^2_\pm$. The averaging period is 2.5 $ \ {\mu}$m. The smoothed results are the final upper and lower limits of the reconstructed mode profile.  Finally, we calculate final reconstructed profile as the mean of the upper and lower limits. The final reconstructed intensity profile is presented in Fig. 4(b). In Fig. 4(b), we see the intensity profile is narrower than the wavelength shift and it has multiple peaks in the center part.  Wiggles outside the center part of the mode are due to Gibbs oscillation \cite{Arfken}. 

The disorder-induced mode profile in Fig. 4(b) is localized mainly from around  $x=12 \ \mu$m to around  $x=38 \ \mu$m. The envelope function has a complicated profile with several local maxima, which are well resolved by our method. Although the spatial extent of the retrieved mode is around 28 $\mu$m which is fairly large compared to modes of high $Q$ photonic crystal cavity such as L3 cavity \cite{noda2002},  thanks to the high $Q$, the value of $Q/V$ as the gauge of Purcell factor \cite{Purcell} of this resonance is similar to the value reported in Ref. 23. It has to be pointed here that the measured $Q$ here is the loaded $Q$ which means two leakage channels contribute here, out-of-plane scattering and loading to the waveguides. From the depth of the resonance dip from our reflection measurement,  we estimate that more than 90$\%$ of the loss is from the out-of-plane scattering for this resonance. 
 
\section{Numerical Validation}
In order to validate our method against a reference mode which is a known disorder-induced resonance, we performed a direct simulation of our experiments.
\begin{figure}[H]%\vspace{-0.25cm}
\centerline{\includegraphics[width=0.9\columnwidth]{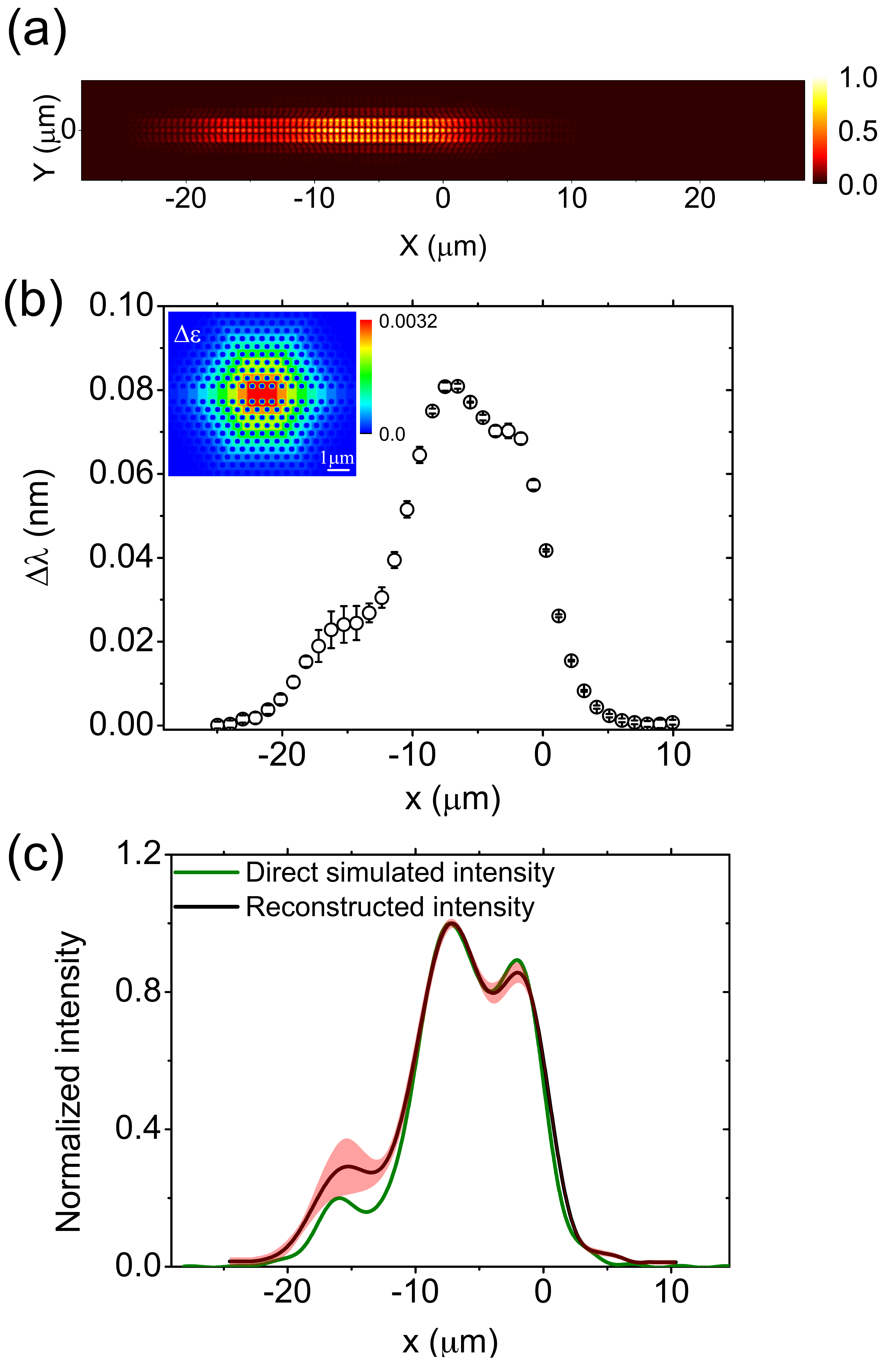}}
\caption{Numerical validation of the mode reconstruction method. (a) Amplitude of the $y$ component of the electric field $|E_y|$ of a simulated mode. (b) Black circles: wavelength shift versus pump spot position. The effect of the pump is simulated by the relative change of the dielectric constant distribution as shown in the top left. (c)  Green curve: directly simulated intensity projected along the $x$ axis. Black curve: reconstructed profile from simulated measurement.}
\end{figure}
The direct simulations are done in two dimensional (2D) finite difference time domain simulations \cite{Oskooi2010}. 2D simulations with an effective index method are valid in a short frequency range for membrane structures \cite{qiu2002}, also they have been used to estimate the mode volume of disordered-induced resonances of experimental realistic photonic crystal waveguides \cite{Henri2010}. Therefore, performing 2D simulation is sufficient to validate our method. 

In our simulation, between the perfect matching layers (PML) and photonic crystal there are dielectric padding layers with refractive index 3.17. In simulations, the lattice constant $a$ of the structure and the radius of the holes are uniformly randomly distributed within the range of $485.00 \pm 4.85$ nm and $121.2500 \pm 1.2125$ nm respectively. This disorder level is the fabrication accuracy of our sample. The length of the waveguide is 106 $a$. For one single realization, in order to obtain the field patterns of disorder-induced modes we follow the following procedure. Firstly, multiple sources with broad linewidth are randomly placed in the waveguide to excite the disorder induced localized modes. Meanwhile, multiple detectors are placed randomly in the waveguide to anaylze the field evolution.  In such a way, we obtain the resonance frequencies of the disorder-induced modes. Secondly, we place multiple sources with narrow linewidth in waveguides in a random way. The sources are spectrally matched to only one of the resonances extracted before. In such a way, we obtain the field pattern of the disorder-induced modes. A disorder-induced localized mode in one random realization with wavelength 2172.855 nm is shown in Fig. 5(a). From the field pattern, we see the simulated disorder-induced localized mode has a periodical oscillation very similar to the waveguide mode, except it is localized rather than extended through the waveguide. 

We create a profile of $\Delta \epsilon(x,y)$ to simulate the change due to the pump,  which is shown in the top left of Fig. 5(b).  It is a discretized Gaussian function with sharp stop. The width of the Gaussian function is 3.5 $\mu$m. The total width of the truncated function is 10.2 $\mu $m. Care has been taken to avoid overlapping boundaries as they lead to unwanted changes in FDTD simulations that are equivalent to extra disorder in the subpixel averaging \cite{FarjadpourRo06}. The pump scan in the experiment is simulated by changing the center of $\Delta \epsilon$. The shift in wavelength versus the pump position of the simulated experiment is shown in Fig. 5(b).  The fine features in Fig. 5(a) are lost because of the width of the simulated pump beam. From Fig. 5(b), we can only obtain the information of where the mode is localized.  By performing the same procedure as for the experimental data, we deconvolve the data in Fig. 5(b) with the cross section of $\Delta \epsilon$.  The result is shown in Fig. 5 (c). The reconstructed intensity has 3 resolvable peaks.  We also plot the integrated intensity along $y$ axis of the mode which is the reference mode in Fig. 5 (c). We see the reconstructed intensity is very similar to the reference mode. It is noticed that there is an overestimation of the amplitude around position $x=- 16 \  \mu$m. This is because of the interaction of a nearby mode with wavelength of 2171.437 nm.  When the pump is moved around $x=- 16 \ \mu$m, the mode with short wavelength has a larger response than the mode we want to retrieve. This causes a weak hybridization of the modes.  As a consequence, the mode we want to retrieve experiences a bigger shift than expected from the first order perturbation theory  (Eq. 1).  Nevertheless, the features and positions of the local peaks are very well resolved.

\section{Conclusion}
In conclusion, we successfully retrieve the spatial profile of a random localized mode in a PhCWG by performing spatially dependent thermal tuning and deconvolution procedures. A direct simulation of our experiment shows that the deconvolution procedure in our method gives a reconstructed mode profile close to the reference mode. Besides measuring the intensity profile of disorder induced modes in PhCWGs which is a 1D system, our method can be straightforwardly applied to higher dimensional disordered systems \cite{Sarma2014,Vynck2012}.  Our method as an accurate tool to retrieve the mode profiles contributes an essential step towards engineering the functionality of the disordered photonic systems \cite{Riboli2014,Wiersma2013}. 

\section*{Acknowledgments} \label{sec:Acknowledgement}
The authors thank Ravitej Uppu and Willem L. Vos for helpful discussions and Cornelis Harteveld for technical support. This work is supported by the European Research Council project No. 279248 and Nederlandse Organisatie voor Wetenschappelijk Onderzoek.

\end{document}